\documentclass[10pt]{iopart}
\usepackage{graphicx}
\usepackage{subfigure}
\usepackage{iopams}
\usepackage[square,sort&compress]{natbib}
\newcommand{\scs}{\scriptscriptstyle}

\newcommand{\bmnab}{\mbox{\boldmath$\nabla$}}

\newcommand{\bmr}{\mbox{\boldmath$r$}}
\newcommand{\bmR}{\mbox{\boldmath$R$}}
\newcommand{\xiB}{\mbox{\boldmath$\xi$}}

\def\a{\alpha}
\def\o{\omega}

\def\e{\epsilon}

\def\<{\langle}
\def\>{\rangle}
\long\def\orangebox#1{%
    \newbox\contentbox%
    \newbox\bkgdbox%
    \setbox\contentbox\hbox to \hsize{%
        \vtop{
            \kern\columnsep
            \hbox to \hsize{%
                \kern\columnsep%
                \advance\hsize by -2\columnsep%
                \setlength{\textwidth}{\hsize}%
                \vbox{
                    \parskip=\baselineskip
                    \parindent=0bp
                    #1
                }%
                \kern\columnsep%
            }%
            \kern\columnsep%
        }%
    }%
    \setbox\bkgdbox\vbox{
        \pdfliteral{1.0  0.898   0.749 rg}   
        \hrule width  \wd\contentbox %
               height \ht\contentbox %
               depth  \dp\contentbox
        \pdfliteral{0 0 0 rg}
    }%
    \wd\bkgdbox=0bp%
    \vbox{\hbox to \hsize{\box\bkgdbox\box\contentbox}}%
    \vskip\baselineskip%
}
\long\def\greenbox#1{%
    \newbox\contentbox%
    \newbox\bkgdbox%
    \setbox\contentbox\hbox to \hsize{%
        \vtop{
            \kern\columnsep
            \hbox to \hsize{%
                \kern\columnsep%
                \advance\hsize by -2\columnsep%
                \setlength{\textwidth}{\hsize}%
                \vbox{
                    \parskip=\baselineskip
                    \parindent=0bp
                    #1
                }%
                \kern\columnsep%
            }%
            \kern\columnsep%
        }%
    }%
    \setbox\bkgdbox\vbox{
        \pdfliteral{0.8  0.898   0.851 rg}   
        \hrule width  \wd\contentbox %
               height \ht\contentbox %
               depth  \dp\contentbox
        \pdfliteral{0 0 0 rg}
    }%
    \wd\bkgdbox=0bp%
    \vbox{\hbox to \hsize{\box\bkgdbox\box\contentbox}}%
    \vskip\baselineskip%
}
\long\def\bluebox#1{%
    \newbox\contentbox%
    \newbox\bkgdbox%
    \setbox\contentbox\hbox to \hsize{%
        \vtop{
            \kern\columnsep
            \hbox to \hsize{%
                \kern\columnsep%
                \advance\hsize by -2\columnsep%
                \setlength{\textwidth}{\hsize}%
                \vbox{
                    \parskip=\baselineskip
                    \parindent=0bp
                    #1
                }%
                \kern\columnsep%
            }%
            \kern\columnsep%
        }%
    }%
    \setbox\bkgdbox\vbox{
        \pdfliteral{0.549  0.698  0.914 rg}   
        \hrule width  \wd\contentbox %
               height \ht\contentbox %
               depth  \dp\contentbox
        \pdfliteral{0 0 0 rg}
    }%
    \wd\bkgdbox=0bp%
    \vbox{\hbox to \hsize{\box\bkgdbox\box\contentbox}}%
    \vskip\baselineskip%
}

\newcommand {\Be}{\begin{eqnarray*}}
\newcommand {\Ee} {\end{eqnarray*}}
\newcommand {\bey} {\begin{eqnarray}}
\newcommand {\eey} {\end{eqnarray}}
\newcommand{\bit}{\begin{itemize}}      
\newcommand{\eit}{\end{itemize}}
\newcommand{\bfl}{\begin{flusleft}}
\newcommand{\efl}{\end{flusleft}}
\newcommand{\bfr}{\begin{flushright}}
\newcommand{\bc}{\begin{center}}
\newcommand{\ec}{\end{center}}
\newcommand{\ben}{\begin{enumerate}}    
\newcommand{\een}{\end{enumerate}}

\def\a{\alpha}
\def\o{\omega}

\def\e{\epsilon}

\def\<{\langle}
\def\>{\rangle}

\begin{document}
\title[Nonlinear excitations match correlated motions unveiled by NMR in proteins]
{Nonlinear excitations match correlated motions unveiled by NMR in proteins:
a new perspective on allosteric cross-talk} 

\author{Francesco Piazza}
\address{Universit\'e d'Orl\'eans, Centre de Biophysique Mol\'eculaire, CNRS-UPR4301,
Rue C. Sadron, 45071, Orl\'eans, France}
\ead{Francesco.Piazza@cnrs-orleans.fr}
\begin{abstract}
In this paper we propose a novel theoretical framework for interpreting
long-range dynamical correlations unveiled in proteins through NMR measurements. 
The theoretical rationale relies on the hypothesis that correlated motions 
in proteins may be reconstructed as large-scale, collective modes sustained by 
long-lived  nonlinear vibrations known as discrete breathers (DB) localized at key, hot-spot sites.  
DBs are spatially localized modes, whose nonlinear nature
hinders resonant coupling with the normal modes, thus conferring them 
long lifetimes as compared to normal modes. DBs have been predicted to exist in proteins, 
localized at few {\em hot-spot} residues typically within the stiffest portions of the structure.
We compute DB modes analytically in the framework of the nonlinear network model,
showing that the displacement patterns of many DBs localized at key sites match to a remarkable 
extent the experimentally uncovered correlation blueprint.  The computed 
dispersion relations prove that it is physically possible for 
some of these DBs to be excited out of thermal fluctuations at room temperature.\\
\indent Based on our calculations, we speculate that transient energy 
redistribution among the vibrational modes in a protein might favor
the emergence of DB-like bursts of long-lived energy at hot-spot sites
with lifetimes in the ns range, able to sustain critical, function-encoding correlated motions.
More generally, our calculations  provide a novel quantitative tool
to predict fold-spanning dynamical pathways of correlated residues that 
may be central to allosteric cross-talk in proteins.
\end{abstract}


\smallskip



%

\section{Introduction}
%
%
Proteins are intrinsically dynamic machines. Their function, {\em e.g.} the ability to 
respond allosterically to a a local perturbation such as ligand binding or chemical modification~\cite{Changeux:2005vn}, 
is intimately related to specific correlated vibrations, whose pattern is rooted in their 
native structure~\cite{Kern:2003uq,Gunasekaran:2004dv,Bahar:1999sc,Marques:1995kq,Li:2004zx}.
Although many experimental techniques lend considerable insight in protein 
dynamics~\cite{Parak:2003zr,Rambo:2013ly,Zaccai:2012ve,Bu2011163},
NMR spectroscopy is emerging as an increasingly powerful method  
for the characterization of correlated vibrational patterns involved in protein 
function and, notably, in allosteric intramolecular 
communication~\cite{Akimoto:2013uk,Selvaratnam:2011ur,Kalodimos:2011kx,Fenwick:2011fk,Ishima:2000ys,Das28082009,Das11072008}.\\
%
%
%
%
\indent A large body of work in the paste decades has contributed to highlight the centrality of 
proteins' three-dimensional folds to their function. Protein scaffolds, for example, 
encode how energy flows across specific pathways encompassing certain {\em hot-spot} 
residues at key locations~\cite{Csermely:2010he}, and linking different functional regions, 
but also sometimes portions of structure not apparently associated with function~\cite{Leitner:2008gl,Moritsugu:2000yo}.
As it is well known, low-frequency, collective normal 
modes (NM) provide key fold-encoded patterns for conformational changes associated with functional 
motions, very often in excellent agreement with observed conformational 
changes~\cite{Tama:2001yc,Kim:2002ni,Ma:1997zs,Zheng:2005ub}. 
However, typically these modes are highly damped due to the substantial 
exposure to the solvent of large portions of the protein 
structure~\cite{McCammon:1987fk,Meinhold:2007uq}, suggesting that other, more localized and therefore
more robust higher-frequency vibrations might be implied in protein function.\\
\indent The hypothesis that protein functional dynamics may imply the concerted action of large-scale 
motions sustained by {\em specific} higher-frequency localized vibrations, although not new~\cite{Garcia:1992fk}, 
has been recently put forward~\cite{Hawkins:2006io} in the context of allostery, 
providing an intriguing rationale for the general idea that both slow and 
fast modes in proteins are connected to function~\cite{Yang:2005qz,Cooper:1984cn,Bahar:1998kx}.
As already argued in the early 70s by C. McClare, for example, enzyme functioning 
may well imply non-thermal storage of energy in specific, fold-rooted localized vibrational 
modes, so as to lower the free-energy barriers of chemical reactions 
{\em where it is needed}~\cite{McClare:1972uq,McClare:1972vn,McClare:1975kx}. 
Indeed, many experiments have affirmed the crucial role of specific localized vibrations, 
such as hinge motions, in mediating between faster atomic fluctuations 
and slower functional rearrangements~\cite{Whitford:2008wo,Eisenmesser:2002or,Henzler-Wildman:2007fc}. \\ 
\indent  Along the same lines, many experimental studies suggest  that unusually 
long-lived vibrational modes may be excited in proteins~\cite{Yu:2003fk,Woutersen:2002uq,Xie:2000ys,Xie:2001fr},
reinforcing the idea that specific nonlinear effects may be central to
their functional dynamics~\cite{dOvidio:2005qy,Scott:1992kx,Archilla:2002ws,Kopidakis:2001iw,Garcia:1992fk}.
As a matter of fact, anharmonic effects are known since a  long time to be highly relevant in many dynamical 
processes in proteins~\cite{Levy:82,Go:95,breath-macromol,Straub:00,Yu:2003fk}. 
For example, recent experiments show that the activation of anharmonic modes is 
required for enzymatic activity in Lysozyme~\cite{Roh:2005fk}.\\
%
%
\indent More recently, it has been shown that nonlinear excitations known in many systems as
Discrete Breathers (DB) allow one to cleverly dissect hot-spot sites and
intramolecular signaling pathways of connected residues critical to protein 
functioning~\cite{Piazza:2009a,Piazza:2011dq}.
DBs are time-periodic, spatially localized vibrational modes that are  found generically
in many-body nonlinear systems~\cite{Flach:2008xy} and possess many properties that make them 
interesting in the context of protein dynamics and allosteric communication. 
It has been shown that DBs promote and sustain long-range dynamical cross-talk 
in proteins, mediating energy transfer to distant locations~\cite{Luccioli:2011bh}.
DBs are able to self-stabilize by harvesting energy from  the background, which has also been 
directly related to DB-mediated long-range communication across protein structures~\cite{Piazza:2011dq,Juanico:2007yw}.
%
%
In fact, at variance with topological excitations such as solitons, interactions between DBs or
between a DB and a large vibrational energy fluctuation 
generally cause a substantial flow of energy from the less energetic to the more energetic mode.
Remarkably, this sheer nonlinear phenomenon has been reported to put distant regions 
of protein structures in communication~\footnote{Although it seems that the link between energy harvesting 
and transfer is indeed related to the emergence  of short-lived nonlinear modes (or large energy fluctuations) 
at key {\em passage} sites~\cite{Piazza:2009a} (making up the energy transduction pathway), 
the exact mechanisms underlying such kind of long-range, spontaneous energy-harvesting phenomena is still unclear.} 
following isolated energy kicks at specific 
locations that trigger the spontaneous emergence of a DB at a distant site~\cite{Piazza:2009a}. 
%
Moreover, by construction DB frequencies do not resonate with normal 
modes, which hampers DB-NM resonant transfer and makes DBs robust against perturbations.
DBs are also protected from solvent-mediated instabilities, as their hot spots lie at 
locations typically far from the protein surface~\cite{Juanico:2007yw}. 
For all these reasons, DBs appear as ideal candidates to realize 
specific long-lived, highly-correlated and fold-spanning motions featuring
high resilience to perturbations. It should be stressed that, in view of their nonlinear character, DBs, 
despite being localized in space, are true {\em collective} motions, where all the particles in the system 
vibrate at one and the same frequency~\cite{Flach:2008xy}. This is the deep reason why 
DB-based analyses are able to unveil fold-rooted dynamical pathways of correlated residues over 
the entire scaffold of a protein~\cite{Piazza:2009a}. \\
%
%
\indent NMR spectroscopy has proved an invaluable tool to dissect long-range correlations in 
proteins~\cite{Bruschweiler:1995fk,Bax:1993uq}. Recently, methods based on statistical analysis of NMR chemical 
shift changes~\cite{Akimoto:2013uk,Selvaratnam:2011ur}
and perturbation patterns~\cite{Zhuravleva:2011uq}
have been introduced, showing a great potential of mapping  
extended networks of coupled amino acids involved in intramolecular
signaling pathways~\cite{Vendruscolo:2011fk}.
In particular, a recent NMR study reported residual dipolar couplings 
measurements revealing a rather puzzling long-lived correlated motion
spanning four $\beta$ strands separated by up to 1.5 nm in Ubiquitin~\cite{Fenwick:2011fk},
showing clear signatures on the msec time scale and raising 
intriguing questions as to the nature of the associated vibrational mode(s). 
The observed correlations resulted from large-scale concerted 
conformational rearrangements partly mediated by the hydrogen-bonding network.
The structures of 640 conformers have been deposited in the PDB repository 
following this study (PDB identifier 2KOX), providing a wealth of precious 
structural information. \\
%
\indent In this paper we show that the vibrational pattern of 
several distinct discrete breather modes found in Ubiquitin at specific key residues match 
to an amazing extent the correlated pattern uncovered experimentally.
Furthermore, we show that the energy required for exciting such modes 
could in principle be available through thermal energy fluctuations occurring at key sites, 
typically residing in the stiffest locations of the protein fold.
One the one hand, our results strongly suggest that the correlations observed experimentally in Ref.~\cite{Fenwick:2011fk}
could flag the spontaneous excitation of discrete breathers in Ubiquitin.
More generally, our calculations suggest that DB-based methods may be central to unveil 
sub-structures comprising residues that mediate 
the transmission of allosteric signals in proteins.\\
\indent The paper is organized as follows. First we introduce our coarse-grained 
model of protein dynamics and describe the essentials of our algorithm for computing 
approximate analytic DB solutions. We then illustrate, in the case of a randomly selected conformer,
how DBs localized at specific sites match the experimental correlated pattern.
Finally, we report the results of the analysis performed over the whole ensemble of 
NMR conformers, which confirms that the same results apply to DBs computed in all experimentally 
resolved structures.
%
\section{Analytic DB modes in the nonlinear network model}
We introduced the Nonlinear Network Model (NNM) with the aim of exploring the subtle effects arising in many-body systems
from the interplay of anharmonicity and the lack of translational order ({\em i.e.} the peculiar 
3$D$ protein folds)~\cite{Juanico:2007yw}, for which 
little is known~\cite{Sukhorukov:2001kl,Rasmussen:1999dp,Kopidakis:2000lr} as opposed to nonlinear systems
with  lattice-like translational invariance~\cite{Flach:1998lg}.
In the NNM a given protein is coarse-grained to the level of amino-acids and 
modeled as a nonlinear network of $N$ identical point-like particles of 
mass $M \approx 110$ a.m.u. ({\em i.e.} the average amino-acid
mass) placed at the corresponding $C_{\alpha}$ sites. 
The NNM potential energy reads
\begin{equation}
\label{FPU}
 V=\sum_{i>j} c_{ij}\left[ 
                             \frac{k_{2}}{2} (r_{ij}-R_{ij})^2 +
                             \frac{k_{4}}{4} (r_{ij}-R_{ij})^4 
                            \right]
\end{equation}
where $r_{ij} = |\bmr_{i} - \bmr_{j}|$ is the distance between particles $i$ and $j$ and
$R_{ij} = |\bmR_{i} - \bmR_{j}|$ is the $i-j$ separation in the equilibrium structure  
(in this case one of the NMR conformers).
Here $\bmr_{i}$ and $\bmR_{i}$ denote the instantaneous and equilibrium position vectors of 
particle $i$, respectively. 
The connectivity matrix is simply $c_{ij}=\{1 \ {\rm if} \ R_{ij}\leq R_{c} | \ 0 \ {\rm otherwise}\}$,
where  $R_c$ is a cutoff that identifies the interacting pairs.
The aim being to explore the connection between nonlinearity and the peculiar structural features of the native folds,
only C$_\a$ atoms are taken into account and $k_{2}$ is  assumed to be the same for all interacting pairs. 
Following our previous studies~\cite{Juanico:2007yw}, we take $R_c=$ 10 \AA, and  
fix $k_{2}$ so that the low-frequency part of the linear spectrum match 
the corresponding measured frequencies. This gives $k_{2}= 5$ kcal/mol/\AA$^2$. 
Furthermore, we fix  $k_{4} = 5$ kcal/mol/{\mbox{\normalfont\AA}}$^4$, which corresponds to a rather weak
nonlinearity~\footnote{For $|r_{ij}-R_{ij}|\approx 0.5$ \ {\mbox{\normalfont\AA}} the 
nonlinear-to-linear energy ratio is  $k_{4}|r_{ij}-R_{ij}|^{2}/2k_{2}\simeq \mathcal{O}(10^{-1})$}.
%
%
From a physical point of view, the constants $k_{2}$ and $k_{4}$ can be rationalized in terms of 
an average inter-residue potential of mean force (PMF). For a given pair of amino acids, a PMF for a given 
reaction coordinate (in our case the C$_{\alpha}$-C$_{\alpha}$ distance) can be computed, {\em e.g.}
via Boltzmann inversion from all-atom molecular dynamics simulations. 
The idea here is that $k_{2}$ and $k_{4}$ measure the coefficients of
the second-order and fourth-order terms in a Taylor expansion of such PMF.
It is interesting to note that reaction coordinates associated with side-chain 
side-chain relative positions ({\em e.g.} the distance between centers of mass) are 
likely to yield strongly anharmonic (flat) PMFs. This is a consequence of the strongly 
nonlinear (multimodal) character of side-chain motions as revealed by the dynamics of 
$\chi_{1}$ dihedral angles~\cite{Garcia:1992fk}.\\
\indent In ref.~\cite{Piazza:2008to} we have introduced a theoretical protocol for calculating 
analytically DB modes of given amplitude $A$ at a given site in the framework of the NNM. 
In principle, like all time-periodic functions, a DB solution 
can be decomposed as a Fourier series comprising harmonics of a fundamental 
frequency.  However, The idea is to start from  an {\em ansatz} for the DB consisting of the simplest time-periodic
function modulated by a time-varying amplitude  
\begin{equation}
\label{e:DBans}
\bmr_{m}(t) = \bmR_{m} + A \, \xiB_{m}(t) \cos \omega t
\end{equation}
where we assume that the spatially localized envelope function $\xiB_{m}(t)$ 
varies slowly on the timescale defined by the inverse DB frequency $\omega^{-1}$.
%
%
This fact can be simply pictured as an alternative statement of DB resilience to 
perturbations~\cite{Flach:2008xy}.
Moreover, we also assume that 
$\max_{m} \xiB_{m}(t) \simeq \mathcal{O}(1)$, so that the parameter $A$ sets the 
physical scale for the oscillation amplitude. 
Under these hypotheses, we can substitute ansatz~\eref{e:DBans} in the equations of motion
and expand the forces in power series of $\epsilon_{ij} = A/R_{ij}$
\begin{equation}
\label{e:eqmotexp}
M\ddot{\bmr}_{m} = -\sum_{p=1}^{3} \frac{1}{p!} \left[ \frac{\partial^{p} }{\partial A^{p}} 
                                     \bmnab_{m} V \right]_{A=0} \! \! \!A^{p} + \mathcal{O}(\epsilon^{4})                            
\end{equation}
We then  multiply Eqs.~\eref{e:eqmotexp} by $\cos \omega t$ and average 
over one DB period. Under the slowly-varying 
envelope approximation, we can thus eliminate the time dependence, {\em i.e.}
\begin{equation}
\left\langle \xiB_{m}(t) \cos^{p} \omega t \right\rangle_{\omega} \approx
\bar{\xiB}_{m} \left\langle \cos^{p} \omega t \right\rangle_{\omega} 
\end{equation}
where $\langle f \rangle_{\omega} = \omega/2\pi\int_{0}^{2\pi/\omega} f(t)\,dt$ denotes the
time average.  
Neglecting by the same token the second time derivatives of the functions $\xiB_{m}(t)$,
we finally map the original set of differential equations  onto a 
nonlinear  system of $3N$ algebraic equations,
whose unknowns are the time-averaged envelope patterns $\bar{\xiB}_{m}$ (normalized to the displacement 
of the central particle)
and the breather frequency. A DB mode with given amplitude $A$ is then found 
by numerically solving the algebraic system corresponding to an appropriate 
initial guess. We have found that the local direction of maximum stiffness
provides an excellent initial guess, ensuring  fast convergence  
and allowing one to investigate the parameter regions where a DB mode 
localized at a given site can be found~\cite{Piazza:2008to}. 
%
\section{Nonlinear correlated motions in Ubiquitin}
%
A general structural analysis proves useful as a start,
as this will provide an instructive reading frame for the results illustrated in 
the rest of the paper. 
The PDB file labelled 2KOX contains 640 
possible conformers of ubiquitin. Thus, in principle a NNM can be constructed starting from each of those
structures and analyzed separately. 
We have shown that few DBs localized at specific key sites, usually within stiff locations, 
appear to be far more stable and robust than all other DBs~\footnote{Different measures of 
local stiffness can be adopted (see Ref. \cite{Juanico:2007yw} for a possible definition), 
but the substance of this observation does not change. }. 
Such DBs arise as nonlinear continuations of high-frequency normal modes, 
which are  spatially localized too, and
invariably feature gap-less excitation spectra at variance with DBs localized at generic sites~\cite{Piazza:2008to}.
This means that we can gain valuable insight by examining the patterns of hot-spot sites highlighted 
by high-frequency NMs within the structural network of the protein over the whole NMR ensemble.\\
%
%
%
%
%
%
\begin{figure}[t!]
\vspace*{.05in} 
\centering
\includegraphics[width=11 truecm,clip]{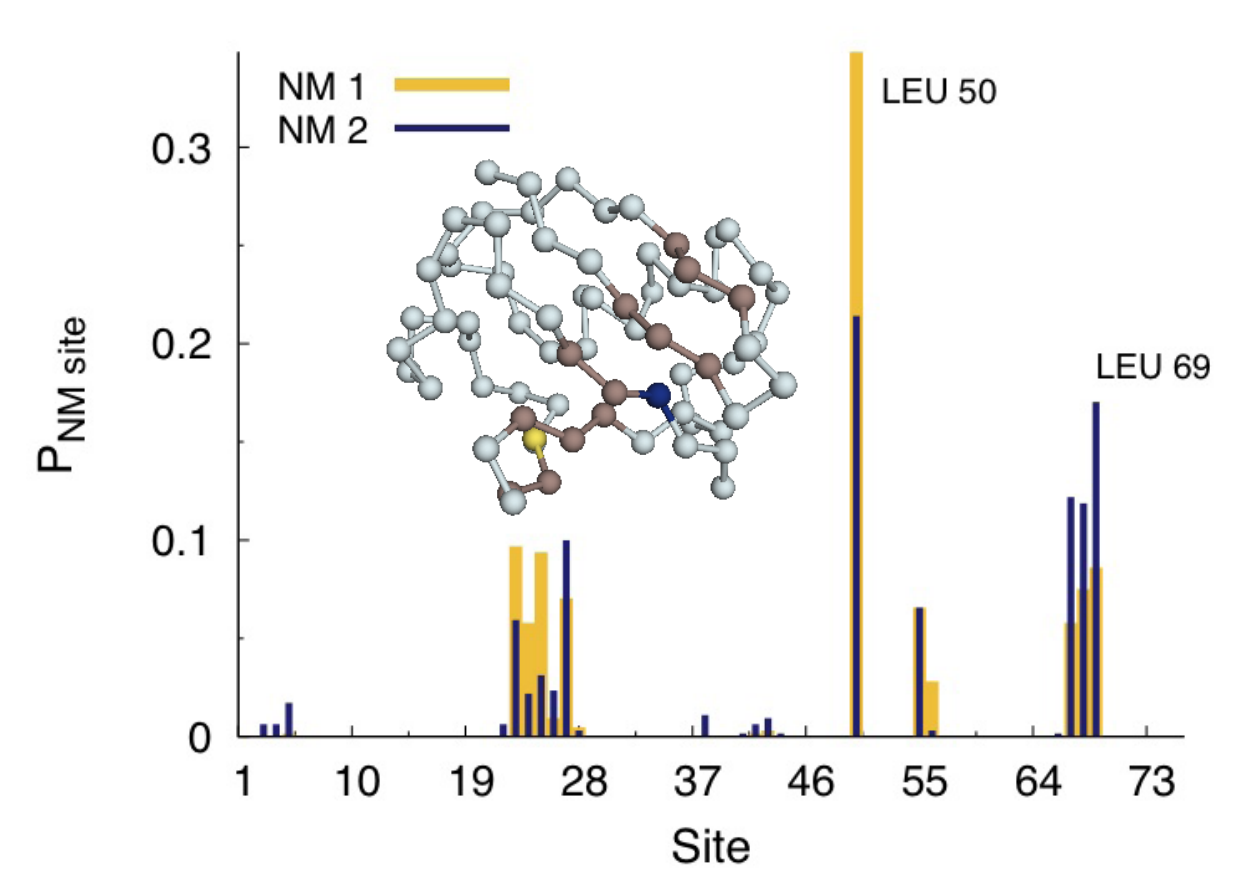}
\caption{\label{f:conn-NM12} (Color online)
Probability  that a residue be the NM site ({\em i.e.} the particle with the largest amplitude) 
of the highest- or second-highest-frequency normal mode.
The cartoon shows a C$_{\alpha}$-coarse-grained structures of Ubiquitin.
The region spanning the $\beta$ strands 
where the correlated motion is detected in the NMR experiments is highlighted in brown.
Residues LEU 50 and LEU 69 are shown as yellow and dark blue spheres, respectively.
The cutoff used to calculate the connectivity matrices is $R_{c}=10$ \AA.} 
\end{figure}
%
\begin{figure*}[b!]
\vspace*{.05in} 
\centering
\subfigure[]{
\includegraphics[width=6.3 truecm,clip]{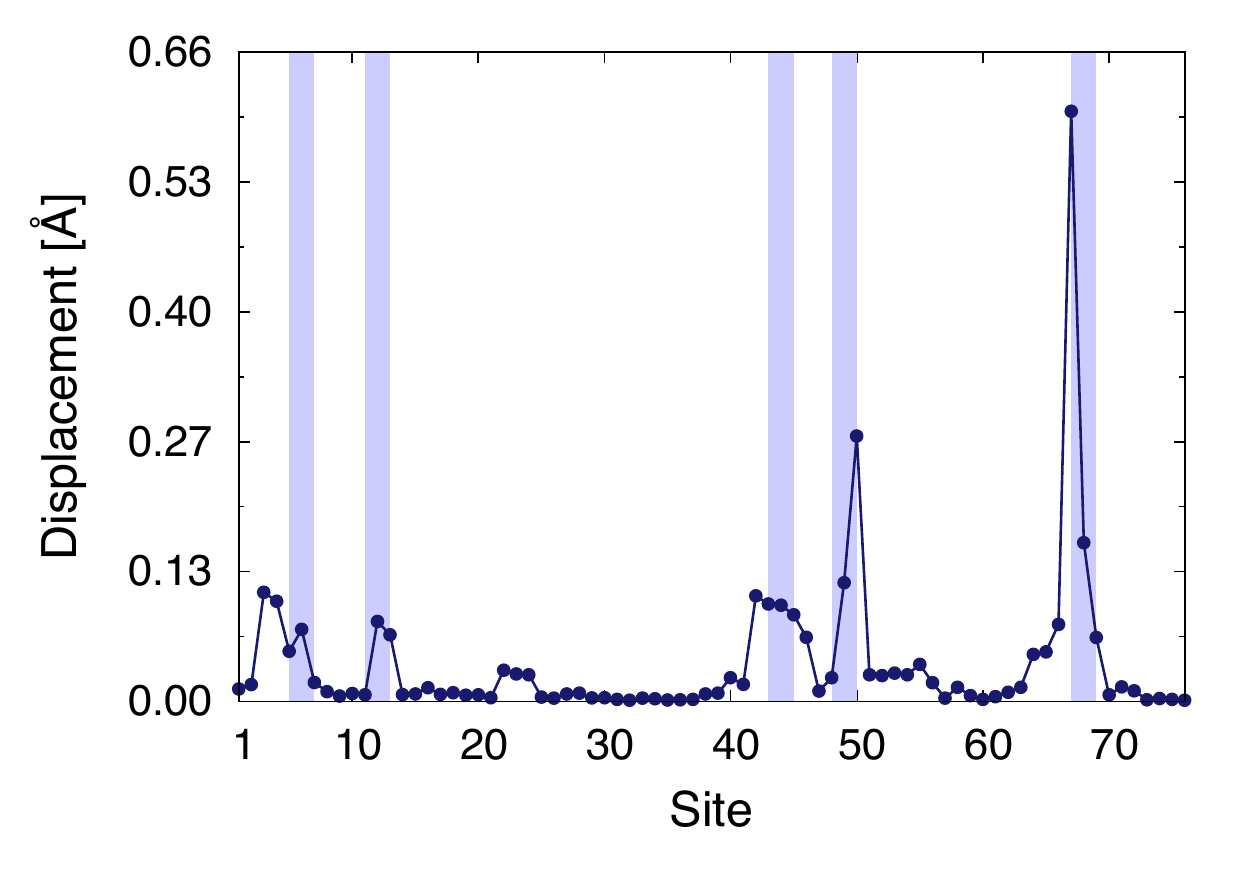}}
\subfigure[]{
\includegraphics[width=6.3 truecm,clip]{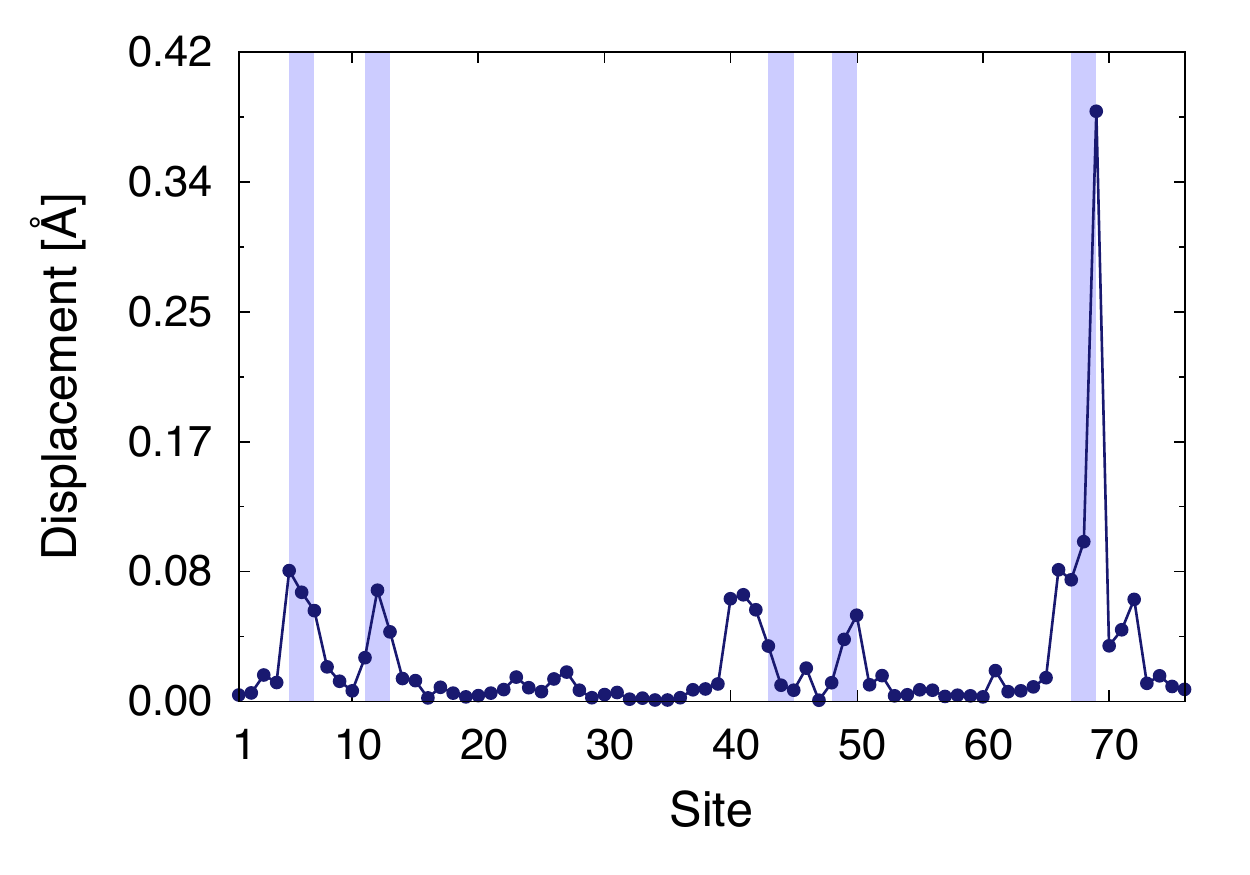}}
\caption{\label{f:0400} (Color online) Displacement patterns of DBs in the 400$^{{\rm th}}$ conformer of
the NMR-resolved Ubiquitin structure. 
%
%
The residues whose correlated motion has been revealed by the NMR measurements 
from ref.~\cite{Fenwick:2011fk} (see~\ref{e:subsetA}) lie within the light blue bands. 
(a) DB at site LEU 67,  $\o_{\scs B}=90.1$ cm$^{-1}$,  $E_{\scs B}=21.4$  kcal/mol.
(b) DB at site LEU 69,  $\o_{\scs B}=86.2$ cm$^{-1}$,  $E_{\scs B}=7.7$  kcal/mol.
Here $E_{\scs B}$ and $\omega_{\scs B}$ denote
the potential energy associated with the displacement field of the DB 
and its frequency,  respectively.
}
\end{figure*}
%
\indent For each conformer, we construct a network as indicated above and compute the 
normal modes, {\em i.e.} the eigenvectors of the Hessian matrix
of the total potential energy $V$~\eref{FPU}.  
Fig.~\ref{f:conn-NM12} reports the pattern of occurrence of the {\em NM sites} of the first two
high-frequency NMs, {\em i.e.} the particles 
whose amplitude of vibration in the mode is largest.
It is apparent that a few sites stand out over the whole ensemble, which identify 
rigid, hinge-like locations~\cite{Juanico:2007yw}. These sites  lie within the stiffest 
regions and are thus subject to small-amplitude fluctuations within the ensemble of
conformers. Overall, hot-spot sites are identified around LYS 27, LEU 50 and LEU 69.
Interestingly, according to NMR measurements reported in Ref.~\cite{Fenwick:2011fk}, 
the last two sites appear to participate to the correlated motion uncovered in the experiment.
More precisely, the latter identify the following subset of residues (see also cartoon in Fig.~\ref{f:conn-NM12})
%
%
\begin{equation}
\label{e:subsetA}
  A =  \left\{
  \begin{array}{l l}
     {\rm VAL}\ \ \ 5 -  {\rm THR} \,7, & {\rm LYS}\,11 - {\rm ILE}\,13, \\
     {\rm LEU}\,43 - {\rm PHE}\,45, & {\rm LYS}\,48 - {\rm LEU}\,50, \\
     {\rm LEU}\,67 - {\rm LEU}\,69
  \end{array}
\right\}
\end{equation}
Our analysis over the ensemble of conformers clearly singles out three different regions as possible 
localization hot-spots for nonlinear localized modes. Thus, the question naturally arises as to what 
is the correlated pattern associated with DBs localized at these special locations.\\
\indent Figures~\ref{f:0400} (a) and (b) show the theoretical 
displacement patterns of two DB modes centered at LEU 67 and LEU 69. These have been calculated 
within the NNM starting from the equilibrium structure corresponding to a randomly picked 
conformer within the NMR ensemble (equivalent pattern maps are obtained by selecting other conformers). 
It can be clearly appreciated that such DB modes could be regarded 
as plausible realizations of the experimentally detected motions, as the displacement 
patterns match to a remarkable extent the experimental observations.\\
%
\begin{figure}[t!] 
\vspace*{.05in}
\centering
\includegraphics[width=10 truecm,clip]{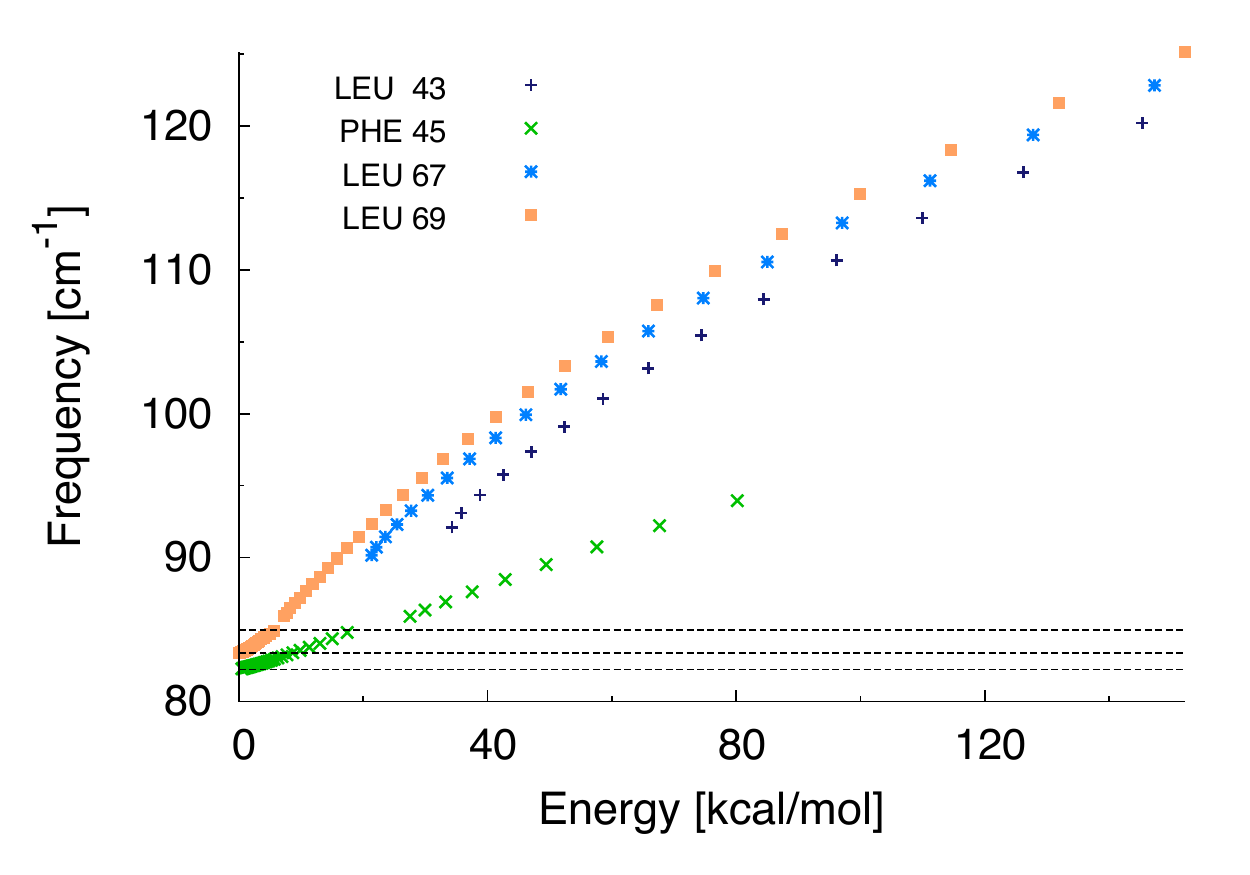}
\caption{\label{f:WE400} (Color online) Frequency vs energy for different Discrete Breathers 
computed in the 400-th conformer. The horizontal dashed lines mark the frequencies of the 
first edge normal modes.}
\end{figure}
%
%
\indent As a general fact, the frequency and energy ranges where DBs  can 
be found in a protein are broad and strongly site-dependent~\cite{Piazza:2008to}. 
%
The dispersion curves shown in 
Fig~\ref{f:WE400} make clear that pattern-matching DBs such as those illustrated in Fig.~\ref{f:0400} 
can be excited at energies as low 
as 5-10 kcal/mol at selected sites with frequency above the linear spectrum. This places
such modes  among the most robust ones, as resonances with NMs are only possible 
with higher harmonics of linear modes. 
%
To be more precise, pattern-matching DBs such as those centered at PHE 45 and LEU 69 
happen to fall within the special class of {\em zero-gap} modes,
that is, they can be excited at arbitrary low energies, and can by all means be regarded 
as analytical continuations of band-edge normal modes that are stabilized through non-linear 
mechanisms~\footnote{As a general feature, in the low-energy limit
zero-gap DBs approach a given high-frequency band-edge normal mode. 
In the analyzed conformer, DBs at LEU 69 and PHE 45 approach 
the 2$^{\rm nd}$  and 3$^{\rm rd}$ NMs (from the band edge), respectively.}. 
Remarkably, this also means that they are among the most stable DBs -- the more energy is 
injected  the more  localized and resilient they become~\cite{Piazza:2008to}.
%
At low energies, such modes are in fact {\em intra-band} breathers, {\em e.g.} their 
frequencies fall within the gap between two successive normal modes. This 
is made possible by the {\em discrete} nature (finite number of particles) of the 
protein and constitutes one of the most distinctive features of localized nonlinear 
modes in discrete systems lacking translational symmetry~\cite{Piazza:2008to}.
%
\\
\indent As a general fact concerning breathers in protein structures~\cite{Piazza:2008to}, the majority of sites 
does not host zero-gap DBs, meaning that it can be exceedingly hard to excite a DB at a generic location,
depending on the associated energy threshold. On the contrary, zero-gap DBs can only be found at very 
few special sites, usually lying within the stiffest regions of the structure~\cite{Piazza:2008to}. 
It is thus tempting to speculate that a particular 
biological relevance can be attached to the patterns of selected zero-gap DBs, which are likely to be 
excited spontaneously at 300 K~\footnote{Of course, this does not mean that the vibrational pattern 
of all zero-gap DBs that may exist in a given protein must have a special biological 
significance.}.  Following Ref.~\cite{Piazza:2011dq}, the average waiting time
between spontaneously occurring uncorrelated energy fluctuations of magnitude $\epsilon$
can be estimated as
\begin{equation}
\label{e:tmed}
\bar{\tau}_{\rm th} \simeq \frac{1}{2\gamma} \sqrt{\frac{\pi}{\beta \e}} \, e^{\beta \e}
\end{equation}
where $\beta = 1/k_{\scs B}T$ and 
$\gamma \simeq 1$ ps$^{-1}$ is a typical damping coefficient specifying the decorrelation 
time scale of atomic tumbling.
For fluctuations in the range $4\div5$ kcal/mol at $T = 300$ K, one gets $\bar{\tau}_{\rm th} \simeq 0.1 \div 1$ nsec. 
This means that  it should be possible to observe the signature of a DB excited out of thermal fluctuations  
(at least) at sites LEU 69 and PHE 45. Therefore, the possibility that DB excitation 
may explain the correlated motions observed experimentally in Ref.~\cite{Fenwick:2011fk}
appears physically realistic.
%
%
%

%
%
\section{Discrete Breathers in the NMR ensemble}
%
%

%
\begin{figure*}[t!] 
\vspace*{.05in}
\centering
\subfigure[]{
\includegraphics[width=10 truecm,clip]{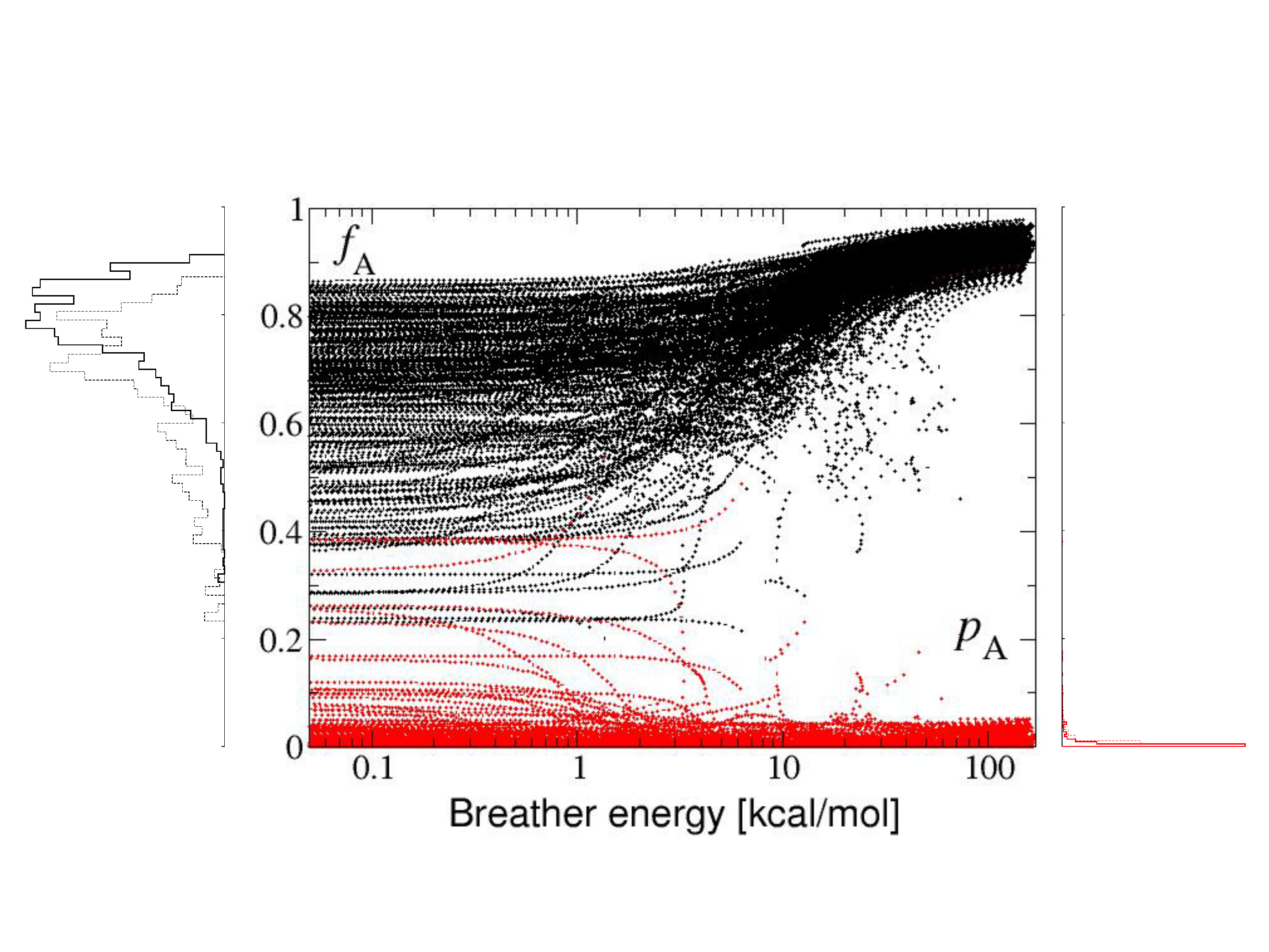}}
\subfigure[]{
\includegraphics[width=10 truecm,clip]{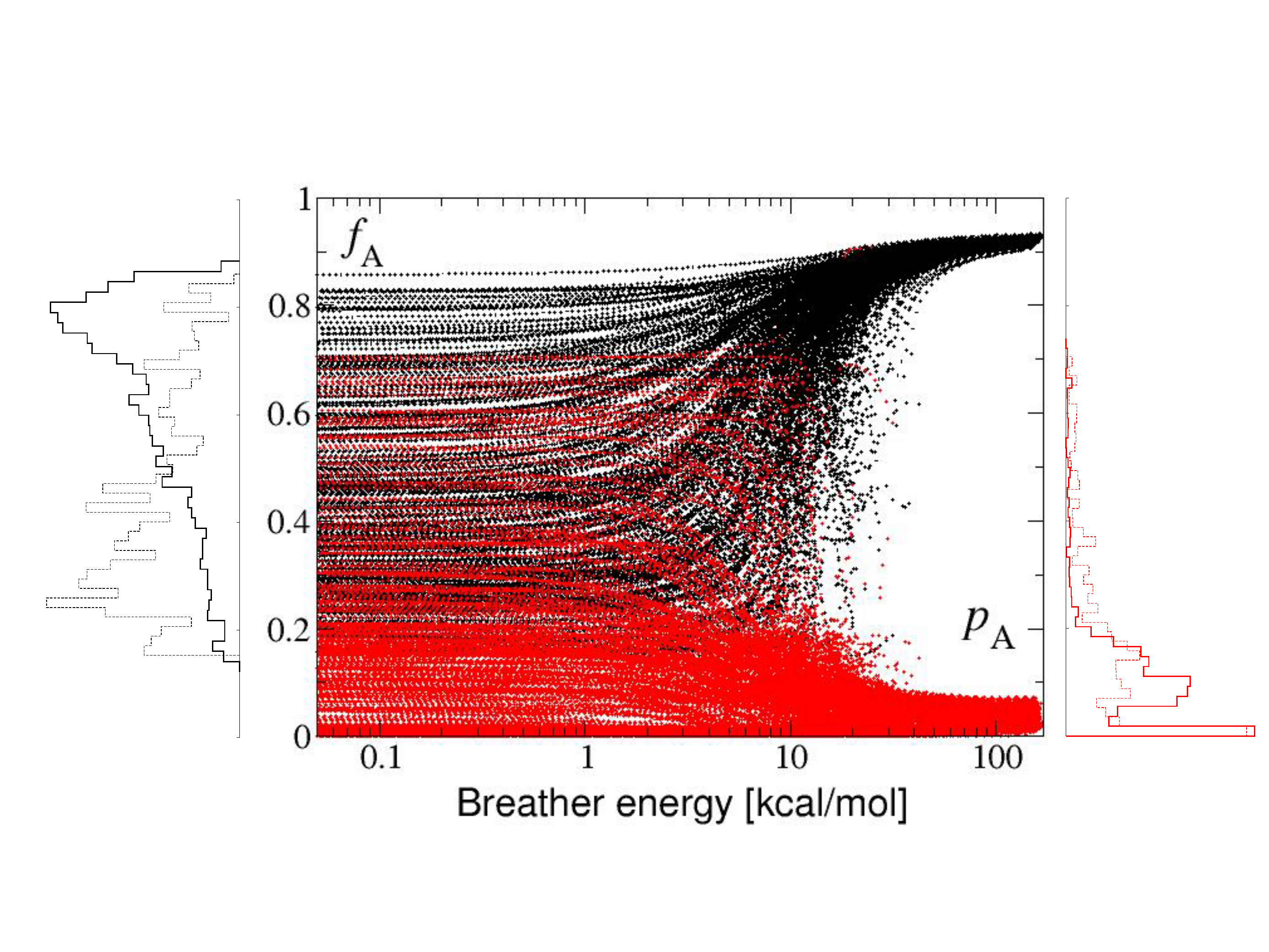}}
\caption{\label{f:fAall50-69} (Color online) Fraction of DB pattern covering 
the experimental correlated motion versus energy for a DB mode centered at LEU 69 (a) and LEU 50 (b).
All DBs found in the entire database of 640 conformers are shown.
The statistical significance of the vibrational pattern matching is illustrated through the 
probability $p_{A}$ (red points), Eq.~\eref{e:pA} (see text).
The side histograms show the distributions of $f_{A}$ and $p_{A}$ over the whole database 
for energies between 7 and 10 kcal/mol (thick solid lines) and for energies
between 0.1 and 1 kcal/mol (thin dashed lines).}
\end{figure*}

\begin{figure*}[t!] 
\vspace*{.05in}
\centering
\subfigure[]{
\includegraphics[width=9 truecm,clip]{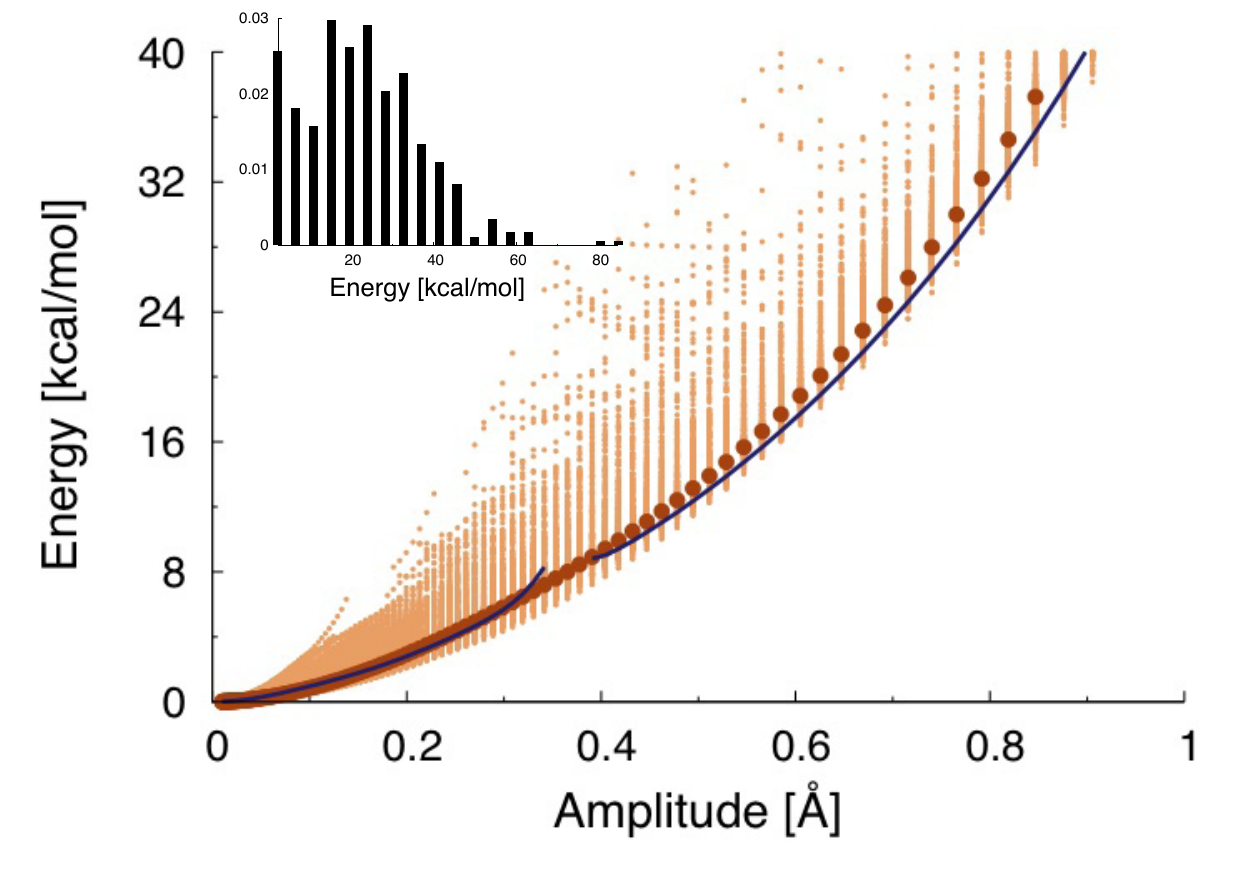}}
\subfigure[]{
\includegraphics[width=9 truecm,clip]{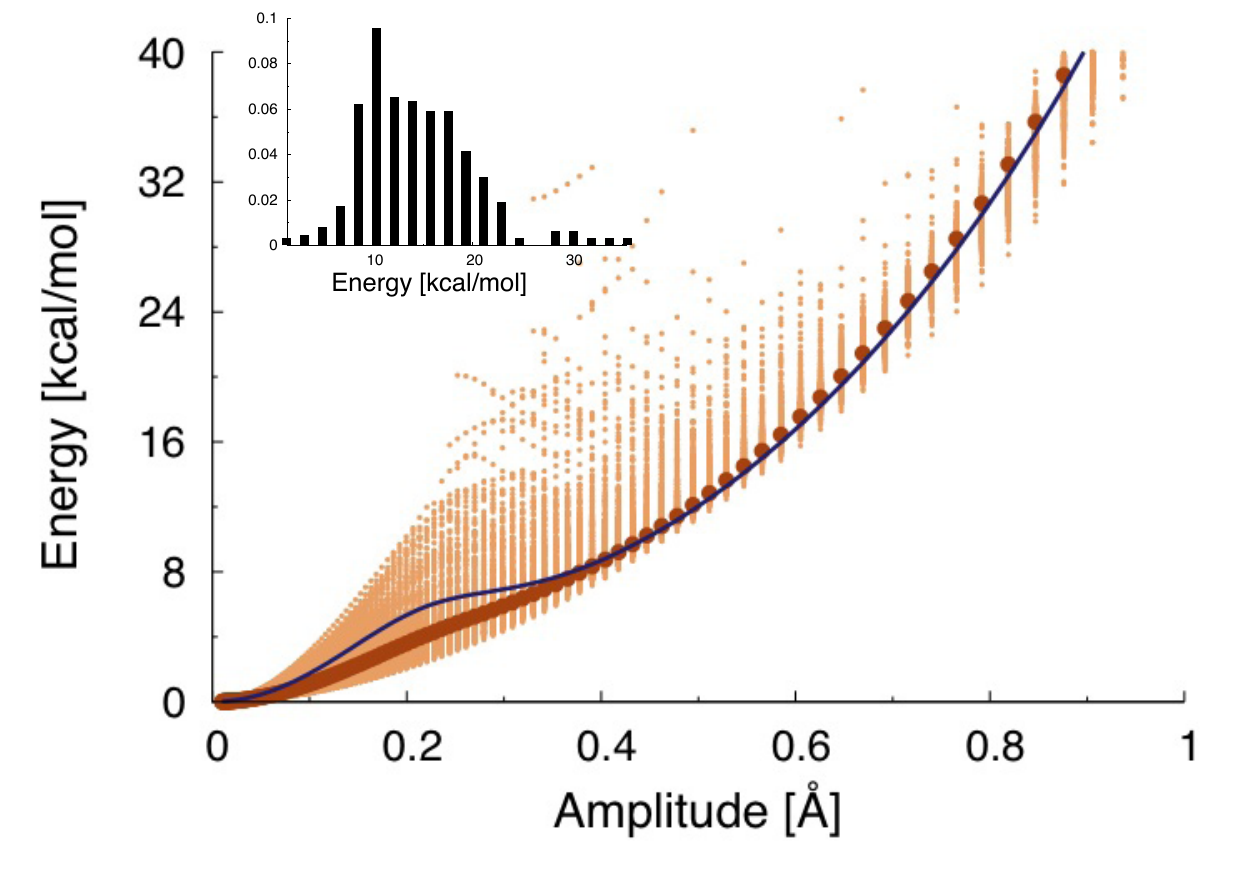}}
\caption{\label{f:WE} (Color online) Amplitude versus energy for Discrete Breathers 
computed in all 640 conformers and centered at (a) LEU 69 and (b) LEU 50 (light filled circles).
The dark filled circles represent the corresponding average curve. The solid 
line refers to the DB computed in the {\em average} structure computed over the NMR ensemble of 
conformers. The insets show the histograms of the energy gaps calculated over the whole ensemble.}
\end{figure*}

We have seen that for a randomly selected conformer
the vibrational patterns of DB modes centered at specific sites 
match to a surprising extent the correlated motion found experimentally. 
However, our calculations take a single conformer as the equilibrium structure for 
constructing the NNM. Therefore, it is necessary to  
investigate whether this is an isolated property displayed by DB modes in a
few special conformers, or rather it reflects a general property of DB
solutions over the ensemble of (experimentally determined) allowed conformations. 
We note that part of the answer is already known, as the few key sites
where the interesting DBs appear lie within stiff regions. Therefore, they 
should retain similar spatial arrangements of their neighborhoods over the 
ensemble (see again Fig.~\ref{f:conn-NM12}). This should guarantee that similar 
DBs should be found at the same hinge-like locations in all the conformers.\\
\indent To answer the above question, we introduce the following energy-dependent indicator  
for a specific DB  whose vibrational pattern is $\xiB_{i}, i=1,2,\dots,N$
\begin{equation}
\label{e:fA}
f_{A} = \sum_{i\in A} |\xiB_{i}|^{2} /\sum_{i=1}^{N} |\xiB_{i}|^{2}
\end{equation}
where $A$ is the subset of residues that participate to the
long-lived experimental motion specified in~(\ref{e:subsetA}).
By construction, the quantity $f_{A}$ measures to what extent 
the pattern of a given DB (with a given energy) in a given conformer matches the experimentally 
highlighted structural correlation.  It should be noted that $f_{A}$ only measures the {\em geographical}
overlap, ignoring by construction 
the specific directions of the vibrational patterns to be compared, as this information is
unfortunately not available from the experiments.\\
\indent Of course, it could be objected that there exists a given probability that 
{\em whatever} pattern match a given region comprising five non-overlapping segments 
of the protein scaffold. For this reason, we have also calculated a $p$--value  
associated with such null hypothesis, {\em i.e.} a measure of statistical significance associated 
with the measured $f_{A}$ values. To this aim, the following quantity has been 
also computed at all values of energies for all conformers
\begin{equation}
\label{e:pA}
p_{A} = \int_{f_{A}}^{1} \mathcal{P}(f) \, df
\end{equation}
where the probability distribution $\mathcal{P}(f)$ has been reconstructed by 
generating a large number of random partitions, each consisting of a random permutation of 
five segments of fixed length (the same lengths as the segments in $A$) centered at as many random 
locations. The quantity $p_{A}$ is  simply the probability that a given match 
score larger than the one actually observed would be observed for a random partition 
with the same structure as $A$. Thus, $p_{A}$ provides an estimate of the rejection 
probability associated with the corresponding $f_{A}$ value.   
The results of these analyses are shown in Figs.~\ref{f:fAall50-69}.\\
\indent As a first global observation, it can be appreciated that the 
fraction of sizable pattern-matching scores is large over the whole 
ensemble of conformers.
Furthermore, while at low energies DBs in different conformers display 
substantially variable scores, above 10 kcal/mol 
all DBs display the same large value of $f_{A}$.
As the pattern-matching score of DBs increases, it can be seen that this is so 
with increasing statistical confidence, as the corresponding rejection probability 
drops to zero. The case of the DB centered at LEU 69 provides a rather clear 
demonstration of this effect. \\
\indent It is interesting to remark that in some cases the fingerprint of the experimental 
correlated pattern seems to be 
present {\em in nuce} already at the harmonic level. This can be clearly appreciated 
by comparing the histograms of $f_{A}$ values for DBs with energies between 0.1
and 1 kcal/mol and between 7 and 10 kcal/mol. 
In the case of DBs localized at LEU 69, for example (Fig.~\ref{f:fAall50-69} (a)), nonlinearity 
manifestly causes a pre-existing, low-energy signature to become sharper. As the  DB becomes more energetic, 
its pattern captures to an increasing extent the experimental correlated motion. In this 
case, one may recognize nonlinear focussing of a pre-existent, fold-encoded vibrational pattern.\\
\indent However, as it is seen from  Fig.~\ref{f:fAall50-69} (b), nonlinearity is also able to promote
a localized mode capturing the experimental pattern at intermediate and
high energies without it showing significant traces in the harmonic regime. At variance with 
DBs localized at LEU 69, DBs centered at LEU 50
change substantially from low to intermediate energies, increasingly focussing their 
vibrational pattern within the experimental region $A$.
In this case, a {\em dynamical rearrangement} occurs through sheer nonlinear effects,
as a DB changes markedly its pattern to match the experimental vibration to an increasing extent 
when its energy builds up. This result demonstrates in a clear fashion that our nonlinear
analysis is able to dig up information unavailable at the NM level. \\
\indent Altogether, the physical properties of the same DBs in different conformers show to be highly consistent.
This is demonstrated by the energy vs amplitude relations displayed in Fig.~\ref{f:WE},
which makes clear that the mean curves calculated by  averaging 
the  corresponding single-conformer relations over the ensemble are in excellent agreement with the properties of the 
{\em average DB}, {\em i.e.} the DB computed in the ensemble-averaged structure. 
Furthermore, we observe that the same gap-less DBs are present in a sizable fraction of the conformers (as {\em e.g.} 
in the case of the DB at LEU 69 whose pattern is shown in Fig.~\ref{f:0400}).
Considering for example the DBs at LEU 69 and LEU 50, we find that 
39 \% and 46 \% of the DBs over the ensemble, respectively, display a vanishing gap.
The average excitation threshold for the rest of the breathers is (LEU 69) $\Delta E = 25.8 \pm 0.6$ 
kcal/mol  and   (LEU 50) $\Delta E = 14.5 \pm 0.3$ 
kcal/mol (see also insets in Fig.~\ref{f:WE}). All in all, it seems that it is easier to excite DBs whose pattern 
bears less resemblance to those of high-frequency normal modes.

%

\section{Conclusions\label{s:conc}}
In this paper we show that the puzzling features of a long-lived, highly correlated motion 
found in Ubiquitin through NMR measurements~\cite{Fenwick:2011fk} match to a surprising extent 
the vibrational pattern of discrete breather modes centered at specific hot-spot 
sites. Our calculations show  that such nonlinear modes could be excited spontaneously 
out of thermal fluctuations at room temperature.
While this could be the first demonstration of the experimental detection of 
DB-like vibrations in a protein,  it is at the same time a powerful demonstration of the ability of
our nonlinear analysis~\cite{Juanico:2007yw} to predict relevant structure-spanning 
dynamical structures that may be central to the transmission of allosteric signals 
across protein scaffolds~\cite{Selvaratnam:2011ur}.\\
\indent The emerging picture is that persistent  
concentration of energy along specific patterns could play a pivotal role in directing conformational changes 
at a higher level. One may speculate that protein folds might encode for the repeated excitation of  
long-lived bursts of higher-than thermal energy localized at specific hot spots located within stiff regions. 
Therefore, pattern-matching large-scale correlations embodied by collective modes
could be effectively sustained by such long-lived nonlinear  vibrations,  that tend to self-focus
where the nodes of the extended modes are. We note that this picture is intriguingly remindful of the 
interplay between nonlinear localized and delocalized motions found by Garcia in atomistic molecular dynamics trajectories 
of a small protein in the early 1990s~\cite{Garcia:1992fk}.  \\
\indent In this scenario, already evoked in the context of enzyme functioning~\cite{Yang:2005qz} and 
allosteric behavior~\cite{Hawkins:2006io}, the node/hot spot pattern complementarity 
of specific slow and fast modes would lie at the core of protein functional dynamics.
%
%
Specific, fold-encoded large-scale motions could be stabilized through nonlinear effects,
thus causing the protein to privilege specific functionally relevant fluctuations, such as those 
of a binding pocket, hinged at some hot-spot site(s), that intermittently but steadily breaths open
or analogous large-scale motions involving cross-talk between hinged domains. 
This picture is ultimately supported by general properties 
of discrete breathers: DBs are not only known for their
resilience to perturbations, low-damping rate due to  reduced contact with the solvent and hindered resonant energy transfer 
with NMs. Remarkably they display the unique property of self-sustaining by {\em harvesting} energy from 
the background~\cite{Flach:1998lg}, which has been recently demonstrated to also occur in proteins~\cite{Piazza:2009a}.\\
%
\indent In summary, our study prompts the intriguing hypothesis that unusually long-lived 
DB-like modes might be central to  rationalize how protein folds encode 
intramolecular cross-talk.
As such, DB-based analyses could provide a key computational method to identify unknown 
dynamical structures at the core of allosteric transduction mechanisms in proteins. 
%
\section{Acknowledgements}
The author is deeply thankful to R. Nussinov, P. Csermely, Y.-H. Sanejouand and  
P. De Los Rios for a critical reading of this manuscript and for their most enlightening comments.
The author is also indebted to R. B. Fenwick, L. Orellana and X. Salvatella for illuminating 
discussions concerning the application of the DB-based method to their NMR data.
Finally, the author would like to thank L. Turin for making him aware of the inspiring 
ideas developed in the early 1970s by C. W. F. McClare. The author acknowledges 
financial support from the EU-FP7 project PAPETS (GA 323901).

%
%
\providecommand{\newblock}{}

%
%
\end{document}